\documentclass[12pt]{article}
\usepackage{amssymb,amsmath}
\newcommand{\PSbox}[3]{\mbox{\rule{0in}{#3}\hspace{#2}\includegraphics{#1}}}

\catcode `\@=11
\def\@cite#1#2{#1\if@tempswa , #2\fi}
\catcode `\@=12

\newtheorem{prop}{Proposition}%
\newtheorem{lm}{Lemma}%
\newcommand{\qed}{\hbox{\rule[-2pt]{3pt}{6pt}}}

\begin{document} 
\title{The free energies of six-vertex models \\
and the $n$-equivalence relation. \\} 
\author{Kazuhiko MINAMI
\thanks{Graduate School of Mathematics, Nagoya University, Nagoya, 464-8602, JAPAN, 
minami@math.nagoya-u.ac.jp}}
\date{7/21/2006}
\maketitle
 
\abstract{ 
The free energies of six-vertex models 
on general domain $D$ 
with various boundary conditions 
are investigated with the use of the n-equivalence relation 
which classifies the thermodynamic limit properties. 
It is derived 
that the free energy of the six-vertex model on the rectangle is unique 
in the limit (height, width) $\to(\infty, \infty)$. 
It is derived  
that the free energies of the model on $D$ are classified 
through the densities of left/down arrows on the boundary. 
Specifically 
the free energy 
is identical to that obtained 
by Lieb and Sutherland with the cyclic boundary condition 
when the densities are both equal to $1/2$. 
This fact explains several results 
already obtained through the transfer matrix calculations. 
The relation to the domino tiling (or dimer, or matching) problems is also noted. }

\newpage

\section{Introduction\label{intro}}
\setcounter{equation}{0}

The six-vertex model is a solvable lattice model, 
usually introduced on the rectangle 
and considered with the use of the Bethe anzats method or the Yang-Baxter relation. 
The model was first solved 
by Lieb[\cite{L1}-\cite{L3}] and generally by Sutherland[\cite{S}], 
both assuming the cyclic boundary condition 
in the horizontal and the vertical directions.  
In this paper $f_{\rm LS}$ denotes 
the free energy obtained by Sutherland. 

We have, on the other hand, an example[\cite{KZJ}] 
where the free energy is exactly obtained 
with another specific boundary condition. 
In this case the free energy is expressed in terms of the elementary functions, 
and thus it is apparently different from $f_{\rm LS}$. 
In the six-vertex model, 
the boundary condition is relevant even in the thermodynamic limit. 
This fact seems to be unusual 
comparing to other lattice models such as the Ising models. 

In this paper 
we consider the free energies of the six-vertex models 
introduced generally on domain $D$ with continuous boundary 
and with various boundary conditions. 
The free energies are investigated and classified, 
with the use of the $n$-equivalence relation, 
and our study includes the cases 
where the transfer matrix method cannnot be directly applied. 
The main result of this paper is proposition \ref{thD} 
which states that 
the density of down arrows and that of left arrows on the boundary 
determine the free energy of the system on $D$. 
The free energy is identical to $f_{\rm LS}$ 
when the two densities are both equal to $1/2$. 
This result also means that the free energy is still intensive 
even if the boundary effect remains relevant in the thermodynamic limit. 

Section \ref{review} 
is a short summary on the six-vertex model 
and the corresponding transfer matrix treatment. 
In section \ref{domain}, we introduce the domain $D$ 
and, in section \ref{neq}, 
introduce an equivalence relation of boundary conditions 
called the $n$-equivalence[\cite{note}]. 
Two boundary conditions yield the identical free energy 
if they are $n$-equivalent. 
This $n$-equivalence is a generalization of the concept of boundary condition, 
classifies the infinite limit properties, 
and also corresponds to the irreducibility of the transfer matrix. 
In section \ref{results}, 
it is derived, 
with the use of the $n$-equivalence,  
that the free energy of the six-vertex model on the rectangle $R$ 
with $w$ columns and $h$ rows 
is unique in the thermodynamic limit $(w, h)\to(\infty, \infty)$, 
specifically independent of the ratio $w/h$, 
independent of the order of two limits $w\to\infty$ and $h\to\infty$. 
The six-vertex model on a cylinder and a rectangle 
are considered and finally we obtain proposition 5. 

There exist several exact calculations which yield $f_{\rm LS}$ 
with various boundary conditions. 
Our results explain why these free energies are equal to $f_{\rm LS}$ and, 
in addition to it,  
can determine the exact free energies of six-vertex models 
which have not yet been solved. 
The results can also be written in terms of the domino tiling language 
and also we can certify that proposition 5 is consistent with the results in this area. 
These facts concerning the relations with other known results are summerized in section 3. 

\section{The six-vertex models on D\label{main}}
\setcounter{equation}{0}
\subsection{The six-vertex model\label{review}}

Let us consider the square lattice and assign an arrow on each bond. 
The arrows are arranged 
such that two arrows come in and the other two go out at each site (the Ice rule). 
Then there exist six types of possible local arrow arrangements as shown in Fig.1. 
In this paper we are going to use the term 'vertex' as a site and four bonds around it. 
Each vertex is assumed to have finite energy. 
The energy is assumed to be unchanged 
by reversing all arrows on the four bonds. 
Then we have three energy parameters 
and hence three types of Boltzmann weights 
$a$, $b$ and $c$ 
assigned to the vertices (see again Fig.1). 
We also introduce the Boltzmann constant $k_{\rm B}$, 
the temperature $T$, $\beta=1/k_{\rm B}T$ 
and the total number of sites $N$. 
The partition function is 
\begin{equation}
Z=\sum_{\rm config.}\prod_{i=1}^{N}e^{-\beta\epsilon_i},
\end{equation}
where $\epsilon_i$ is the energy of the $i$-th vertex 
and $\sum_{\rm config.}$ is taken over all the possible arrow configurations. 
The free energy $f$ is obtained through 
$-\beta f=\lim_{N\rightarrow\infty}N^{-1}\log Z$. 

One can assign a line on each arrow pointing down or left (Fig.1). 
Then each arrow configuration corresponds to a line configuration on the lattice.  
The Ice rule corresponds to the restriction that 
each line begins from a bond on the boundary, 
continue until it reaches another bond on the boundary, 
and that the lines do not intersect each other. 

Let us consider a rectangle $R$ with $w$ columns and $h$ rows. 
Assume that $h$ and $w$ are even. 
Let $\eta\equiv\{x_1, \ldots, x_m\}$ be a line configuration 
with $m$ lines on a row of vertical bonds in $R$, 
and let $\eta'\equiv\{x'_1, \ldots, x'_m\}$ be that on the row below. 
The symbol $\{x_1, x_2, \ldots,  x_m\}$ denotes that 
there is a line on each $x_k$-th bond $(k=1, \ldots, m)$ 
and otherwise there is not. 
The $(\eta',\eta)$-element of the transfer matrix $V$ is introduced as 
\begin{eqnarray}
V_{\eta'\eta}&\equiv&\langle x'_1, \ldots, x'_m|V|x_1, \ldots, x_m\rangle
\nonumber\\
&=&\sum_{\rm config.} \prod_{k=1}^w e^{-\beta\epsilon_k}, 
\end{eqnarray}
where $|x_1, \ldots, x_m\rangle$ is the state 
corresponding to $\eta\equiv\{x_1, \ldots, x_m\}$, 
$\epsilon_k$ $(k=1, \ldots, w)$ is the energy of the $k$-th vertex on the row 
between the two rows of vertical bonds 
and here the sum $\sum_{\rm config.}$ 
is taken over all the possible line configurations 
with fixed $\eta$ and $\eta'$. 

When we identify the bond at the right end with that at the left end on each row  
(we call this the cyclic boundary condition in the horizontal direction), 
the transfer matrix is identical for all rows. 
If the boundary condition is also cyclic in the vertical direction 
the partition function $Z$ is written as 
\begin{equation}
Z={\rm tr}\; V^h=\sum_{i} \lambda_i^h
\sim\lambda_1^h\;\; (h\rightarrow\infty), 
\end{equation}
where $\lambda_i$'s are the eigenvalues of $V$ 
and $\lambda_1\geq|\lambda_i|$ for all $i$. 

Following the notations in [\cite{B}] 
let us introduce $\Delta=(a^2+b^2-c^2)/2ab$. 
The transfer matrix is block diagonalized 
according to the number of lines $m$. 
The maximum eigenvalue of the transfer matrix 
lies in the block element 
with $m=0$ or $w$ ($\Delta> 1$), with $m=w/2$ ($\Delta < 1$). 
When $\Delta>1$, 
we have two types of frozen phase  
where specific line configurations are dominant. 
In this case 
the free energy is a constant 
and all the arguments in this paper become trivial. 
We thus concentrate on the case with $\Delta < 1$.  

\subsection{Domain $D$\label{domain}}

We introduce the domain $D$. 
Let us consider a continuous and closed line 
$\gamma(t)=(x(t), y(t))\; (0\leq t \leq 1)$ 
which satisfies 
$\gamma(t_1)\neq\gamma(t_2)$ if $t_1\neq t_2$ except $\gamma(0)=\gamma(1)$. 
Let us assume that the sites are on the points $(n_1a_1, n_2a_2)$, 
where $n_1$ and $n_2$ are integers, $a_1$ and $a_2$ are the lattice spacings.  

The sites inside $\gamma$ belong to $D$. 
The sites on the line $\gamma$ can be suitably defined 
to belong or not to belong to $D$. 
The vertices belong to $D$ when the corresponding sites belong to $D$. 
Each row and column in $D$ is assumed to be simply connected. 

The bonds/vertices in $D$ are called the boundary bonds/vertices 
when they have non-zero intersection with $\gamma$. 
The sites of the boundary vertices are called the boundary sites. 
It is assumed that 
the number of vertices on the boundary divided by the total number of sites 
vanish in the thermodynamic limit. 

We will take the following limit:  
fix the line $\gamma$ and take the thermodynamic limit $a_1, a_2\to 0$. 
This corresponds to taking the limit $w , h\to\infty$
where $w$ and $h$ are the number of columns and rows in $D$. 

\subsection{The n-equivalence\label{neq}}
Now we introduce the $n$-equivalence[\cite{note}]. 
Assume that each site and bond takes 
one of a finite number of states. 
Let us consider the set of sites 
which lie in $D$ and can be reached from the boundary sites 
by $n$ steps ($n$ bonds) at minimum; 
we call these sites the $n$-boundary sites. 
Consider the set of bonds 
between $(n-1)$- and $n$-boundary sites, 
and call them the $n$-boundary bonds. 
At last let us consider the $n$-boundary sites together with the $n$-boundary bonds 
and call them the $n$-boundary. 
Configurations on the $n$-boundary are called $n$-boundary configurations.  
Let $\{ \Gamma_i\}$ be the set of all the possible configurations 
on the $n$-boundary 
with a boundary condition $\Gamma$ on the actually boundary of $D$. 
Two boundary conditions $\Gamma$ and $\Gamma'$ 
are called $n$-equivalent  
when $\{ \Gamma_i\}=\{ \Gamma'_i\}$ as a set of $n$-boundary configurations. 

\begin{prop}\label{nequiv}
Fix a sequence of finite lattices $\{D_N\}$, 
where $N$ is the number of sites 
and $D_N$ approaches to the thermodynamic limit as $N\to\infty$. 
Suppose that the boundary conditions $\Gamma$ and $\Gamma'$ 
are $n$-equivalent, for each $D_N$, 
with $n=o(N/N')$ where $N'$ is the number of boundary sites. 
Then the two free energies with $\Gamma$ and $\Gamma'$ are identical 
in the thermodynamic limit. 
\end{prop}
Proof:
Let us write the partition function of the system on $D_N$ as $Z=\sum_i B_iZ_i$. 
The factor $Z_i$ is the partition function from the variables 
inside the $n$-boundary with a fixed $n$-boundary configuration $\Gamma_i$. 
The factor $B_i$ is the contribution from the other variables 
with the boundary condition $\Gamma$ and the fixed $\Gamma_i$. 
The factor $B_iZ_i$ is thus the partition function 
with the $\Gamma$ and the $\Gamma_i$. 
From the assumption that $n=o(N/N')$, 
we obtain  
$\log Z_i =-\beta Nf_i+o(N)$ and $\log B_i =o(N)$. 
The index $i$ runs from $1$ to $i_{\rm max}$, 
where $i_{\rm max}$ is the number of permitted configurations on the $n$-boundary 
and satisfies $i_{\rm max}\leq O(r^{N'})\;$ 
where $r$ is a constant.  
Then we obtain
\begin{eqnarray}
\frac{1}{N}\log Z&=&\frac{1}{N}\log(\sum_iB_iZ_i)
\nonumber\\
&=&\frac{1}{N}
\log B_1Z_1[1+\sum_{i=2}^{i_1}\frac{B_i}{B_1}\frac{Z_i}{Z_1}
             +\sum_{i=i_1+1}^{i_{\rm max}}\frac{B_i}{B_1}\frac{Z_i}{Z_1}]
\nonumber\\
&=&\frac{1}{N}
\log B_1Z_1[1+\sum_{i=2}^{i_1}e^{o(N)}
             +\sum_{i=i_1+1}^{i_{\rm max}}e^{-\beta N(f_i-f_1)+o(N)}]
\nonumber\\
&\rightarrow&-\beta f_1  \hspace{1.2cm} (N\rightarrow\infty),
\end{eqnarray} 
where $f_i=f_1 \; (1\leq i\leq i_1)$, $f_1 < f_i \; (i_1+1 \leq i \leq i_{\rm max})$.
\qed
\newline

Note that what is important is the set of possible configurations, 
we do not need think about the number of ways 
in which each configuration is realized. 

This $n$-equivalence is a generalization of the concept of boundary condition, 
and can be considered in other lattice models such as the 19-vertex model, 
and also can be introduced in stochastic processes[\cite{note}]. 
All the boundary conditions are 1-equivalent, for example,  
in the case of Ising models with finite interactions, 
because the Ising spin states are independent of their nearest-neighbors. 
It follows that the free energies of 
the Ising models are independent of their boundary conditions. 

The equivalence can be introduced in a more generalized form. 
We can consider a boundary condition $\Gamma$ on a subset of the boundary, 
and introduce the corresponding $n$-boundary and the corresponding $n$-equivalences. 
In propositions \ref{limit} and \ref{propalt}
we concentrate on one and two of the four edges of the rectangle $R$ 
and consider the corresponding $n$-equivalences. 

\subsection{Results\label{results}}

\begin{lm}\label{m-cyc}
Consider a rectangle $R$ 
and assume the cyclic boundary condition in the horizontal direction. 
Then all the line configurations with $m$ lines 
on the upper edge (the first row of vertical bonds) of $R$ 
are $2m$-equivalent to each other. 
\end{lm}
Proof:
Let  $\{x_1, x_2, \ldots,  x_m\}$, $x_i< x_{i+1}$, 
be a line configuration on the first row of $R$. 
Beginning from a line arrangement 
$\{x_1, x_2, x_3, x_4, x_5\}=\{2, 3, 5, 6, 9\}$  
as shown in Fig.2(a), for example, 
one can introduce the shift of lines 
$\{2, 3, 5, 6, 9\}\rightarrow \{1, 2, 3, 5, 6\} \rightarrow \{1, 2, 3, 4, 5\}$. 
When we generally have 
\newline
$\{1, 2, \ldots, i, x_j, x_{j+1}, \ldots, x_{m-k}\}$, $i+1<x_j$, after $k$-th step, 
one can introduce the line arrangement 
$\{1, 2, \ldots, i, i+1, x_j, \ldots, x_{m-(k+1)}\}$ as the next one. 
We need $m-l$ steps 
for the shift $\{x_1, x_2, \ldots, x_m\}\rightarrow \{1, 2, \ldots, m\}$
if $x_l \leq m < x_{l+1}$, 
and this means $\{x_1, x_2, \ldots, x_m\}$ 
is at most $m$-equivalent to $\{1, 2, \ldots, m\}$. 
Then all the line configurations $\{x'_1, x'_2, \ldots, x'_m\}$ 
are again possible on the $2m$-boundary. 
\qed
\newline

Let us consider line configurations, 
on the left and right edges,  
with $m_2$ lines in the first $h_p$ bonds on the boundary, 
and $m_2$ lines in the next $h_p$ bonds, and so on. 
In this case one can introduce the line density $\rho_2$ 
on the boundary as $\rho_2=m_2/h_p$. 

\begin{lm}\label{m-fix}
Suppose that the boundary configurations 
on the right and the left edges of $R$ are identical and fixed 
with the density $\rho_2=m_2/h_p$ where $0<\rho_2<1$. 
Then all the line configurations with $m$ lines 
on the upper edge of $R$ 
are $\bar{m}$-equivalent where $\bar{m}=\alpha m+\alpha'$ ($\alpha$, $\alpha'$ are constant). 
\end{lm}
Proof:
Let us introduce the shift of lines 
$\{1, 2, \ldots, i, x_j, x_{j+1}, \ldots, x_{m-k}\}$$\to$
$\{1, 2, \ldots, i, i+1, x_j, \ldots, x_{m-(k+1)}\}$, $i+1 < x_j$,  
on each row without lines on the right and left boundary bonds, 
and  
$\{x'_1, x'_2, \ldots, x'_m\}$$\to$$\{x'_m, x'_1, \ldots, x'_{m-1}\}$ 
on each row with lines on the boundary bonds, 
as shown in Fig.2(b). 
The line configuration will be arranged as 
$\{x_1, x_2, \ldots, x_m\}\to$$\{1, 2, \ldots, m\}$ 
within $m'$ steps where $m'=[m/(h_p-m_2)]h_p+h_p$. 

Next we consider 
an arbitrary line configuration $\{x''_1, x''_2, \ldots, x''_m\}$, 
$x''_i < x''_{i+1}$, and the shift 
$\{1, 2, \ldots, m\}$$\to$$\{x''_1, x''_2, \ldots, x''_m\}$. 
Let us introduce 
$\{x'_1, x'_2, \ldots, x'_m\}$  $\to$ 
$\{x'_1, x'_2, \ldots, x'_m\}$   
on each row without lines on the right and left boundary bonds, 
$\{1, 2, \ldots, i, x''_{i+1}, \ldots, x''_m\}$ $\to$ 
$\{x''_m, 1, 2, \ldots, i-1, x''_i, x''_{i+1}, \ldots, x''_{m-1}\}$, 
as shown in Fig.2(c),  
on each row with lines on the boundary bonds. 
Finally the configuration $\{x''_1, x''_2, \ldots, x''_m\}$ 
is possible on the $\bar{m}$-boundary 
where $\bar{m}\leq m'+m''$ and $m''=[m/m_2]h_p+h_p$. 
\qed 
\newline

Boundary line configurations with which  
no configuration is admitted on the whole of the lattice 
should be excluded from our argument, 
because in this case the system cannot be a six-vertex model. 
We also assume the convergence of the free energy 
in the sequential limits $h\to\infty$ and $w\to\infty$, 
and another sequential limits $w\to\infty$ and $h\to\infty$. 

\begin{prop}\label{limit}
Consider the six-vertex model on a rectangle $R$. 
Assume the cyclic boundary condition in the horizontal direction, 
or otherwise 
assume that the boundary configurations 
on the right and the left edges of $R$ 
are identical, periodic with fixed period $h_p$, 
with line density equal to $\rho_2=m_2/h_p$ 
satisfying $0<\rho_2<1$. 
Assume that $h$ is always a multiple of $h_p$. Then,  
\begin{itemize}
\item[i)]
the transfer matrix of each row, 
or the product of the transfer matrices of sequential $h_p$ rows, 
respectively, 
is block-diagonalized according to the number of lines 
and each block element is irreducible, 
\item[ii)]
the free energy is unique 
in the limit $(w, h)\to(\infty, \infty)$, 
specifically the limit is independent of the order of two limiting procedures 
$w\to\infty$ and $h\to\infty$, 
and also independent of the ratio $w/h$ 
when one take $h\to\infty$ with fixed $w/h$. 
\end{itemize} 
\end{prop}
Proof: 
First let us assume that the boundary condition is cyclic in the horizontal direction. 
The case with the fixed boundary condition can be treated similarly, 
and will be considered at the last of the proof. 
Let $m$ be the number of lines on the upper edge 
(the first row of vertical bonds) of $R$. 
Then there are $m$ lines on every row of $R$ 
because of the line conservation and the boundary condition. 
Thus the transfer matrix $V$ is block-diagonalized according to $m$.  
Let $V_m$ be the block element of $V$ with a fixed line number $m$. 
All the elements of $V_m$ are non-negative 
because they are sums of Boltzmann weights. 
From lemma \ref{m-cyc}, 
all the line configurations with $m$ lines are $2m$-equivalent. 
Hence there always exist allowed line arrangements on the lattice 
for any line configuration with $m$ lines on the upper boundary
(the first row of vertical bonds) 
and for any line configuration with $m$ lines on the $n$-boundary with $n\geq 2m+2m$, 
and thus we find that all the elements of $V_m^{4m}$ are positive. 
(All the elements of $V_m^{2m}$ are already positive, 
which is obvious from the proof of lemma \ref{m-cyc}.) 
Hence we find that $V_m^{4m}$ is irreducible, 
because any matrix is irreducible if 
all of its elements are positive.  
It follows that $V_m$ is irreducible 
because $V_m^{4m}$ cannot be irreducible if $V_m$ is not, 
this proves i).  
Then the Frobenius theorem works and we know the followings. 
There exists an non-degenerate eigenvalue $\lambda_1(w)>0$ 
such that $\lambda_1(w)\geq|\lambda_i(w)|$ 
where $\lambda_i(w)$ $(i\geq 2)$ are the other eigenvalues of $V_m$. 
We also know that 
there exists an eigenvector associated with $\lambda_1(w)$ 
with all the elements being positive, 
i.e. the projections satisfy $\langle x_1, \ldots, x_m|{\rm max}\rangle$$> 0$ 
where $|{\rm max}\rangle$ is the eigenstate 
associated with the maximum eigenvalue of $V_m$. 
These results are valid for every finite $m$. 

The partition function $Z$ with finite $w$ and $h$ 
is written as 
\begin{eqnarray}
Z
&=&\langle x'_1, \ldots, x'_m|V^h|x_1, \ldots, x_m\rangle
\nonumber\\
&=&c_1\lambda_1(w)^{h}+\sum_{i\geq 2} c_i\lambda_i(w)^{h}
\end{eqnarray}
where
$\{x_1, \ldots, x_m\}$ and $\{x'_1, \ldots, x'_m\}$ are the line configurations 
on the upper and the lower edges of $R$, respectively, 
$c_1=
\langle x'_1, \ldots, x'_m|{\rm max}\rangle\langle{\rm max}|x_1, \ldots, x_m\rangle$ 
is positive and independent of $h$  
and here we have assumed $|{\rm max}\rangle$ is normalized. 
(The coefficients $c_i$ are independent of $h$ when $V$ is diagonalizable. 
Otherwise $V$ is expressed in Jordan form, ¡¡ 
$c_i$ $(i\geq 2)$ are asymptotically bounded by polynomials of $h$ with finite degrees, 
and the following argument remains still valid.) 
We will consider the limit of   
\begin{eqnarray}
-\beta f_{h,w}=\frac{1}{hw}\log Z
&=&\frac{1}{w}\log\lambda_1(w)+\frac{1}{h}z'(h,w)
\label{lnZ-hw}
\\
z'(h,w)
&=&\frac{1}{w}\log c_1[\;1+\sum_{i\geq 2}\frac{c_i}{c_1}(\frac{\lambda_i}{\lambda_1})^{h}]. 
\label{lnZ-lambda}
\end{eqnarray}

The factor $z'(h,w)$ is finite when $h$ and $w$ are finite,  
$z'(h,w)$ depends on $h$ only through the second term  
and so remains finite 
in the limit $h\to\infty$ with fixed $w$.
When we take the limit $h\to\infty$ in (\ref{lnZ-hw}), 
the term $(hw)^{-1}\log Z$ is convergent, 
because the term $w^{-1}\log\lambda_1(w)$ is independent of $h$ 
and the facter $h^{-1}z'(h,w)$ converges to zero. 
Taking $w\to\infty$  afterwards, 
one find that $w^{-1}\log\lambda_1(w)$ is convergent 
because $(hw)^{-1}\log Z$ is assumed to be convergent in this limit. 

On the otherhand, 
when one first take the limit $w\to\infty$ in (\ref{lnZ-hw}), 
one find that the factor $z'(h,w)$ is convergent, 
because $(hw)^{-1}\log Z$ is assumed to be convergent 
and now it can be used that $w^{-1}\log\lambda_1(w)$ is also convergent in this limit. 
Therefore $z'(h,w)$ is bounded by a factor $C$ 
which is independent of $w$, 
because convergent series are always bounded. 
Again note that  
$z'(h,w)$ is finite in the limit $h\to\infty$, 
and therefore the factor $C$ can be taken as a constant independent of $w$ and $h$. 

Hence from (\ref{lnZ-hw}) we have found that 
\begin{eqnarray}
|-\beta f_{h,w}-\frac{1}{w}\log\lambda_1(w)|\leq\frac{C}{h}
\label{fhw-ineq}
\end{eqnarray}
which means that 
the convergence of $f_{h,w}$ in the limit $h\to\infty$ is uniform throughout $w$. 
This proves ii) when we consider the known fact that 
the uniformity of the convergence 
yields uniqueness of the limit of double series. 

The last fact is well known  
but we are going to show a proof of it. 
We have shown that 
the convergence in the limit $h\to\infty$ is uniform: 
there exists a number $f_{\infty,w}$ which satisfies that, 
for every $\epsilon>0$ 
there is an integer $h_0(\epsilon)$ which is independent of $w$, 
such that $|f_{h,w}-f_{\infty,w}|<\epsilon$ for all $h\geq h_0(\epsilon)$. 
The free energy is convergent 
in the next limit $w\to\infty$: 
there exists a number $f$ which satisfies that, 
for every $\epsilon>0$ 
there is an integer $w_0(\epsilon)$ 
such that $|f_{\infty,w}-f|<\epsilon$ for all $w\geq w_0(\epsilon)$. 
Then for all $h, w\geq{\rm max}\{h_0(\epsilon), w_0(\epsilon)\}$ we have 
\begin{equation}
|f_{h,w}-f|\leq|f_{h,w}-f_{\infty,w}|+|f_{\infty,w}-f|< 2\epsilon, 
\label{epsd}
\end {equation}
which means 
\begin{equation}
\lim_{(h,w)\to(\infty, \infty)} f_{h,w}=f 
\end{equation}
as a double series. 
We assumed the convergence 
$f_{h,w}\to f_{h,\infty}$ $\;(w\to\infty)$, 
thus taking $w\to\infty$ in (\ref{epsd}) 
one obtains 
$|f_{h,\infty}-f|\leq 2\epsilon$ 
which means 
\begin{equation}
\lim_{h\to\infty}\lim_{w\to\infty} f_{h,w}=f. 
\end{equation}
Taking $h\to\infty$ also in (\ref{epsd}) 
one obtains 
$|f_{\infty,w}-f|\leq 2\epsilon$ 
which means 
\begin{equation}
\lim_{w\to\infty}\lim_{h\to\infty} f_{h,w}=f.
\end{equation}

At last let us consider the case where 
the boundary line configurations on the right and the left edges of $R$ 
are identical, fixed 
and the lines are located periodically with the period $h_p$ on the edges, 
with the line density being $\rho_2=m_2/h_p$ satisfying $0<\rho_2<1$. 
Then one can introduce a transfer matrix $V=V_1V_2\cdots V_{h_p}$ 
where $V_1$, \ldots, $V_{h_p}$ are the transfer matrices 
of sequential $h_p$ rows of $R$, respectively. 
The partition function 
is expressed as a linear combination of $\lambda_i(w)^{h/h_p}$, 
where $\lambda_i(w)$ are now the eigenvalues of $V=V_1V_2\cdots V_{h_p}$. 
Then lemma \ref{m-fix} works and we obtain the same result. 
\qed
\newline

Note that the $2m$-equivalence of line configurations corresponds 
to the irreducibility of the block element $V_m^{4m}$, 
and hence to the irreducibility of $V_m$. 
The matrix $V$ is irreducible 
if the line configurations used as a bases for the matrix representation of $V$ 
are $n$-equivalent to each other for some finite $n$. 
When the boundary is cyclic in the horizontal direction, 
$\lambda_1(w)$ is already known 
and we have $\lim_{w\to\infty}w^{-1}\log\lambda_1(w)=-\beta f_{\rm LS}$. 
Hence the free energy itself is actually obtained in the proof of proposition \ref{limit}. 
Here we show what is obtained from our formula using only the $n$-equivalence 
and the uniform convergence of the free energy. 

\begin{lm}\label{prophalf} 
Consider the six-vertex model on a rectangle $R$. 
Assume the cyclic boundary condition in the horizontal direction, 
and assume that the number of lines on the upper and the lower edges 
are $m=m(w)$. 
Then all the boundary line configurations with the same $m(w)$ 
yield the identical free energy in the limit $(w, h)\to(\infty, \infty)$. 
Specifically we have $f=f_{\rm LS}$ when $m=w/2$. 
\end{lm}
Proof: 
The line configuration on the upper edge is $2m$-equivalent 
to arbitrary line configurations 
in the block element of $V$ with $m$ lines, 
and hence $2m$-equivalent to the cyclic boundary with $m$ lines. 
We take the limit $h\to\infty$ with fixed $w$, 
and obtain the free energy with $(w, h)\to(w, \infty)$. 
The resulted functions are the same 
for all of these fixed and the cyclic boundary conditions with $m$ lines 
on the upper and the lower edges. 
Next taking the limit $w\to\infty$, 
the free energies remain identical. 
In particular, we obtain the maximum eigenvalue of $V$ 
and the known free energy $f_{LS}$ when $m=w/2$. 
\qed

When the system is cyclic in two directions, 
we know that 
the maximum eigenvalue of the the row to row transfer matrix 
lies in the block-element with $w/2$ lines in each row. 
Because of the symmetry of the system, 
the maximum eigenvalue of the the column to column transfer matrix 
also lies in the block element with $h/2$ lines in each column, 
otherwise we have contradictions. 

When the boundary is cyclic in the horizontal direction 
with fixed $h/2$ lines in each column, 
the 'alternate' line configuration with $\rho_2=1/2$ with the period $h_p=2$ 
is possible on the right and left edges, 
hence lemma \ref{m-fix} works, 
and in lemma \ref{prophalf}, $f_{\rm LS}$ appears with the restriction that 
the number of lines is $h/2$ in each column. 
(Note that proposition \ref{nequiv} is valid 
when at least one necessary configuration is possible on the lattice.) 

\begin{lm}\label{prophalf2} 
In lemma \ref{prophalf}, $f_{\rm LS}$ appears 
with the restriction that 
the number of lines is $h/2$ in each column. 
\end{lm}
Let us consider a fixed boundary line configuration 
with a periodic pattern on the boundary of $R$. 
We assume that $w$ and $h$ are multiples of $w_p$ and $h_p$, respectively, 
and the line density is $\rho_1=m_1/w_p$ on the upper and the lower edges, 
$\rho_2=m_2/h_p$ on the left and the right edges. 
The boundary line configuration on the upper edge is identical to that on the lower edge, 
and the configuration on the right edge is identical to that on the left edge. 
The limit will be taken with fixed $w_p$ and $h_p$. 
In the proof of the next proposition 
we do not need the explicit form of $\lambda_1(w)$. 

\begin{lm}\label{propalt}
With these conditions 
the line densities $\rho_1$ and $\rho_2$ 
determine the free energy of the six-vertex model on $R$: 
$f=f(\rho_1, \rho_2)$.
Specifically $f(1/2, 1/2)=f_{\rm LS}$.   
\end{lm}
Proof: 
First assume $0< \rho_1 < 1$ and $0 < \rho_2 < 1$. 
Lemma \ref{m-fix} yields that 
all the line configurations on the upper edge with the density $\rho_1$ 
are $n$-equivalent with some $n$ which depends only on $w$, 
and all the line configurations on the lower edge with the density $\rho_1$ 
are also $n$-equivalent with the same $n$. 
The boundary effect is relevant to the limit 
only through $\rho_1$ when we take $h\to\infty$ with fixed $w$. 
Next taking $w\to\infty$ we obtain the thermodynamic limit. 
This argument is also valid for the right and left edges with the density $\rho_2$ 
taking the limit $w\to\infty$ with fixed $h$ at first, 
and next $h\to\infty$. 
The limit is unique because of the proposition \ref{limit}. 

Consider the case with $\rho_1=\rho_2=1/2$. 
The boundary configurations on the upper and the lower edges 
are $n$-equivalent to the cyclic boundary with $w/2$ lines 
with some $n$ which depends only on $w$. 
Taking $h\to\infty$ with fixed $w$, 
we obtain the limit identical to that 
obtained with the cyclic boundary condition in the vertical direction with $w/2$ lines 
together with fixed boundary configurations with $\rho_2=1/2$ 
on the right and left edges. 
Next taking $w\to\infty$ we obtain the thermodynamic limit, 
and from lemma \ref{prophalf2} and proposition \ref{limit},  
the limit is identical to $f_{\rm LS}$. 
 
When $\rho_2=0$ the free energy is $(1-\rho_1)\epsilon_1+\rho_1\epsilon_2$, 
and when $\rho_1=0$ we have $(1-\rho_2)\epsilon_1+\rho_2\epsilon_2$. 
As for the cases with the densities equal to $1$, 
one can use the fact that $f(\rho_1, \rho_2)=f(1-\rho_1, 1-\rho_2)$ 
which comes from the symmetry of vertex energies. 
\qed
\newline

Let us consider the sequence of vertical boundary bonds on the line $\gamma$. 
Let us assume a fixed boundary line configuration 
with a periodic pattern on this sequence 
with the line density $\rho_1=n_1/w_p$, 
i.e. $n_1$ lines on every $w_p$ vertical bonds on $\gamma$. 
Also assume a fixed and periodic boundary line configuration 
on the sequence of horizontal bonds on $\gamma$ 
with the line density $\rho_2=n_2/h_p$, 
i.e. $n_2$ lines on every $h_p$ horizontal bonds on $\gamma$. 
Assume that 
there exists a line on a horizontal bond on the left edge of $D$ 
if and only if there is a line on the horizontal bond 
of the same row on the right edge of $D$. 
Assume also that 
there exists a line on a vertical bond on the lower edge of $D$ 
if and only if there is a line on the vertical bond 
of the same column on the upper edge of $D$. 
The limit will be taken with fixed $w_p$ and $h_p$. 
With these conditions 
we can derive the following:  

\begin{prop}\label{thD} 
The line densities $\rho_1$ and $\rho_2$ 
determine the free energy of the six-vertex model on $D$: 
$f=f(\rho_1, \rho_2)$.
Specifically we obtain $f(1/2, 1/2)=f_{\rm LS}$.   
\end{prop}
Proof: 
Let $D_0$ be a rectangle of the width $\Delta x$ and the height $\Delta y$ 
sufficiently large satisfying $D\subset D_0$. 
The rectangle $D_0$ 
is divided into small rectangles $R_i$  
of the width $\Delta x$ and the height $\Delta y_i=\Delta h$, 
where the lower edge of $R_i$ coincide with the upper edge of $R_{i+1}$. 
Let $D'=\cup_{i} R'_i$, 
where $R'_i$ is a rectangle 
of the height $\Delta y_i$ and of the width $\Delta x_i$, 
satisfying $R'_i\subset R_i$ and $R'_i\subset D$,  
and satisfying that the number of columns in each $R'_i$ 
is a multiple of $w_p$ 
taking its maximum with the restriction that $R'_i\subset D$.
The sites in $R'_i$ and those on the edges of $R'_i$, if any, 
are assumed to belong to $R'_i$, 
with the exception that the sites on $R'_i\cap R'_{i+1}$ belong to $R'_i$. 
The sum of the line numbers on the upper and the right edges of each $R'_i$ 
is generally equal to the sum of the line numbers 
on the lower and the left edges of $R'_i$, 
because of the line conservation property. 
Assume that the boundary line configuration on each edge of $D'$ 
is identical to the corresponding boundary line configuration of $D$, i.e. 
there exists a line on a horizontal bond on a right edge of $D'$
if and only if there is a line on the same row at the right edge of $D$, 
and so on for other edges. 

We have assumed in each $R'_i$ that 
the number of columns is a multiple of $w_p$. 
In addition to it, 
we assume in each $R'_i$ that 
the number of rows is also a multiple of $h_p$. 
The latter choice corresponds to introducing a sequence $a_2\to 0$ 
where each $a_2$ satisfies $\Delta h=n'h_pa_2$ with $n'=1, 2, \ldots$.  
It is sufficient to consider this specific sequence  
because  
the difference from the remaining cases of $a_2$ vanish as $N\to\infty$ 
together with the ratio $N'/N$ 
where $N'$ is the number of boundary sites.   

Because of the identical line configuration on the right and left edges, 
the number of lines on the upper and the lower edges of $R_i'$ are the same. 
All the possible configurations on the upper edge are $n$-equivalent 
with some $n$ independent of $a_2$,  
all the configurations on the lower edge are also $n$-equivalent, 
and hence they yield the identical free energy in each $R_i'$ 
in the limit $a_2\to 0$. 
Next taking $a_1\to 0$, 
we find that each rectangle $R'_i$ yields $f(\rho_1,\rho_2)$ 
which is the free energy obtained in lemma \ref{propalt}. 
Thus we obtain $f_{D'}=f(\rho_1,\rho_2)$ 
where $f_{D'}$ is the free energy on $D'$. 

The result is independent of the ratio $\Delta h/\Delta x_i$ 
and the ratio can be taken sufficiently small. 
The ratio of the energy contribution, 
the contribution from $D\backslash D'$ over that from $D$, goes to zero 
in the limit $\Delta h\to 0$, 
because the line $\gamma$, which determines the boundary of $D$, is continuous.  
Therefore we obtain $|f_D-f_{D'}|<\epsilon$ for arbitrary positive $\epsilon$. 
\qed
\newline

The result means that the free energy is still additive 
even in the situation 
where the boundary condition remains relevant 
in the thermodynamic limit. 

These results are applications of [\cite{note}] 
to the case of the six-vertex model, 
and give sufficient conditions to have $f_{\rm LS}$. 
It should be noted that 
we didn't need diagonalize sequential transfer matrices 
but it was sufficient for us to consider the configurations 
which are $n$-equivalent, 
for the purpose to classify the thermodynamic limit properties. 

\section{Conclusion\label{conclusion}}
\setcounter{equation}{0}

Our propositions explain several results already obtained 
and also able to determine the exact free energies of six-vertex models 
which have not been solved. 

One can introduce boundary conditions, 
such as the cyclic boundary condition, 
in which various boundary configurations are admitted. 
If our vertex energies and the temperature satisfy $\Delta < 1$, 
and if we can find configurations 
being $n$-equivalent to those with $\rho_1=\rho_2=1/2$, 
then the free energy is $f_{\rm LS}$. 
If we are in the parameter region with $\Delta >1$, 
and if one of the boundary configurations 
with $(\rho_1, \rho_2)=(0, 0)$ or $(1, 1)$ is admitted, 
the system falls in trivial frozen phases. 

Owczarek and Baxter[\cite{OB}] 
solved (by the Bethe ansatz method)
the six-vertex model with the cyclic boundary 
and a 'free' boundary condition, respectively, in two directions. 
Batchelor {\it et al.}[\cite{BBRY}]  
solved (by the Yang-Baxter relation) 
the six-vertex model on the rectangle $R$ 
with a specific boundary condition 
in which they have assumed 
that the arrow at one end of a row points right (left)
if that on the other end of the same row points left (right), 
and also assumed the cyclic boundary condition in the vertical direction. 
One can find that the boundary configuration with the line density equal to $1/2$ 
is realized with the restrictions in both [\cite{OB}] and [\cite{BBRY}], 
and hence it can be directly derived from our results 
that the free energies of these systems are $f_{\rm LS}$. 
Furthermore one can easily construct a bunch of $n$-equivalent cases 
which have not yet been solved 
and are extremely difficult to solve 
directly by the Bethe ansatz method or the Yang-Baxter relation, 
but we now know that the free energies of all of these cases 
should be $f_{\rm LS}$. 

We would like to note that 
equivalences of boundary conditions in the six-vertex model 
is also investigated in [\cite{Wu}] yielding $f_{\rm LS}$. 

One of the most interesting problems related to our results is the domino tiling 
(see [\cite{Propp},\cite{Ken}]). 
The problem is to find the number of possible ways  
to cover a region completely using dominos ($1\times 2$ rectangles). 
It is of course equivalent to find the number of dimer coverings on a given lattice. 

Each configuration of domino is expressed in terms of a height function $h(x)$. 
The correspondence is unique except an overall constant. 
More precisely we introduce $h(x)$ as follows: 
color the squares in a checker-board pattern, 
and $h(x)$ increases one in each unit 
moving anti-clockwise around black squares on the boundary of dominos, 
and decreases one around white squares, as shown in Fig.3(a). 
Not all the regions can be tiled using dominos. 
The necessary and sufficient condition for tilability 
is written in terms of the height function as follows: 
$|h(x)-h(y)|\leq d(x,y)$ for all $x$ and $y$ on the boundary of the region 
where $d(x,y)$ is the minimal number of steps 
moving from $x$ to $y$ with only black squares on its left. 

Kasteleyn[\cite{Kast}] and independently Temperley and Fisher[\cite{TF}] derived 
the number of tilings on the $m\times n$ rectangle 
and obtained that in the thermodynamic limit it behaves $\exp(mnG/\pi)$ 
where $G$ denotes the Catalan's constant 
$G=1/1^2-1/3^2+1/5^2-1/7^2+\cdots$. 
The number of tilings on the Aztec diamond 
(the 'square' rotated $\pi/4$ tiled by horizontal and vertical dominos) 
is also obtained[\cite{Aztec}] exactly as $2^{n(n+1)/2}$ 
where $n$ is the half of the diameter of the region. 
This is completely different from that of the $m\times n$ rectangle. 
We recognize that 
the number of possible ways of tiling strongly depends on the shape of the boundary. 

Cohn, Kenyon and Propp[\cite{CKP}] showed a variational principle 
for the number of tilings: 
assume that the region is tilable and sufficiently 'fat', 
and assume that the slope of the height function 
$(s,t)=(\partial h/\partial x, \partial h/\partial y)$ 
is asymptotically constant on the boundary,  
then the asymptotic number of tilings per domino is a function of $(s,t)$. 

It is known[\cite{KZJ},\cite{FW}] that the number of possible domino tilings per domino 
is equal to the partition function of the six-vertex model 
with $a=b=1$ and $c=\sqrt{2}$, i.e. $\Delta=0$. 
The equivalence is obtained 
from the correspondence of configurations of dominos and vertices 
shown in Fig.3(b). 

The boundary of the $m\times n$ rectangle has constant slope $(0,0)$ and, 
as a six-vertex model, 
line densities are $\rho_1=1$ and $\rho_2=0$ on the upper edge. 
Mixing two boundaries shown in Fig.3(c) periodically, 
we find that 
the line densities vary as $\rho_1=1-\epsilon$ and $\rho_2=0+\epsilon$ 
while the slope remains $(0,0)$. 
Taking $\epsilon=1/2$ we have $\rho_1=\rho_2=1/2$. 
This modification can also be done for other three edges 
and we find, from proposition \ref{thD}, that the number of tilings 
is obtained from the partition function by Lieb and Sutherland 
with $a=b=1$, $c=\sqrt{2}$ that means $\Delta=0$. 
The free energy is (see for example [\cite{B}]) 
\begin{equation}
f_{LS}=-k_{\rm B}T\int_{-\infty}^{+\infty}
\frac{\sinh^2\frac{\pi}{2}x}{2x\sinh\pi x\cosh\frac{\pi}{2}x}dx.
\end{equation}
Counting the residues at $z=i, 3i, 5i, \ldots$ on the imaginary axis, 
we find the Catalan's constant $G$ 
and obtain that $-\beta f=2G/\pi$.  
The factor $2$ corresponds to the fact 
that the number of vertices is equal to the number of dominos 
and twice the number of squares, 
and the result from our proposition \ref{thD} is consistent 
with that previously obtained by Kasteleyn and by Temperley-Fisher.  
The limit $m\to\infty$ and $n\to\infty$ in the domino case is unique, 
which is also consistent with our proposition \ref{limit}.  

The six-vertex model with the domain wall boundary condition[\cite{KZJ}]  
corresponds to the Aztec diamond. 
In this case the boundary line densities do not satisfy our condition, 
the free energy is not indentical to $f_{LS}$. 

Quite recently, S.Sheffield informed me his work[\cite{Sf}] 
in which he derived a variational principle  
for the systems with gradient Gibbs potential. 
His argument is based on the DLR condition, 
equivalent to the variational principle[\cite{CKP}] 
in the case of the domino tiling problems,  
and corresponds to proposition \ref{thD} 
in the case of the six-vertex model 
while the treatments of the thermodynamic limit is different.  
\newline

The author would like to thank Professor V. Korepin and T. Hattori for discussions.

\newpage
\begin{figure}[htbp]
\PSbox{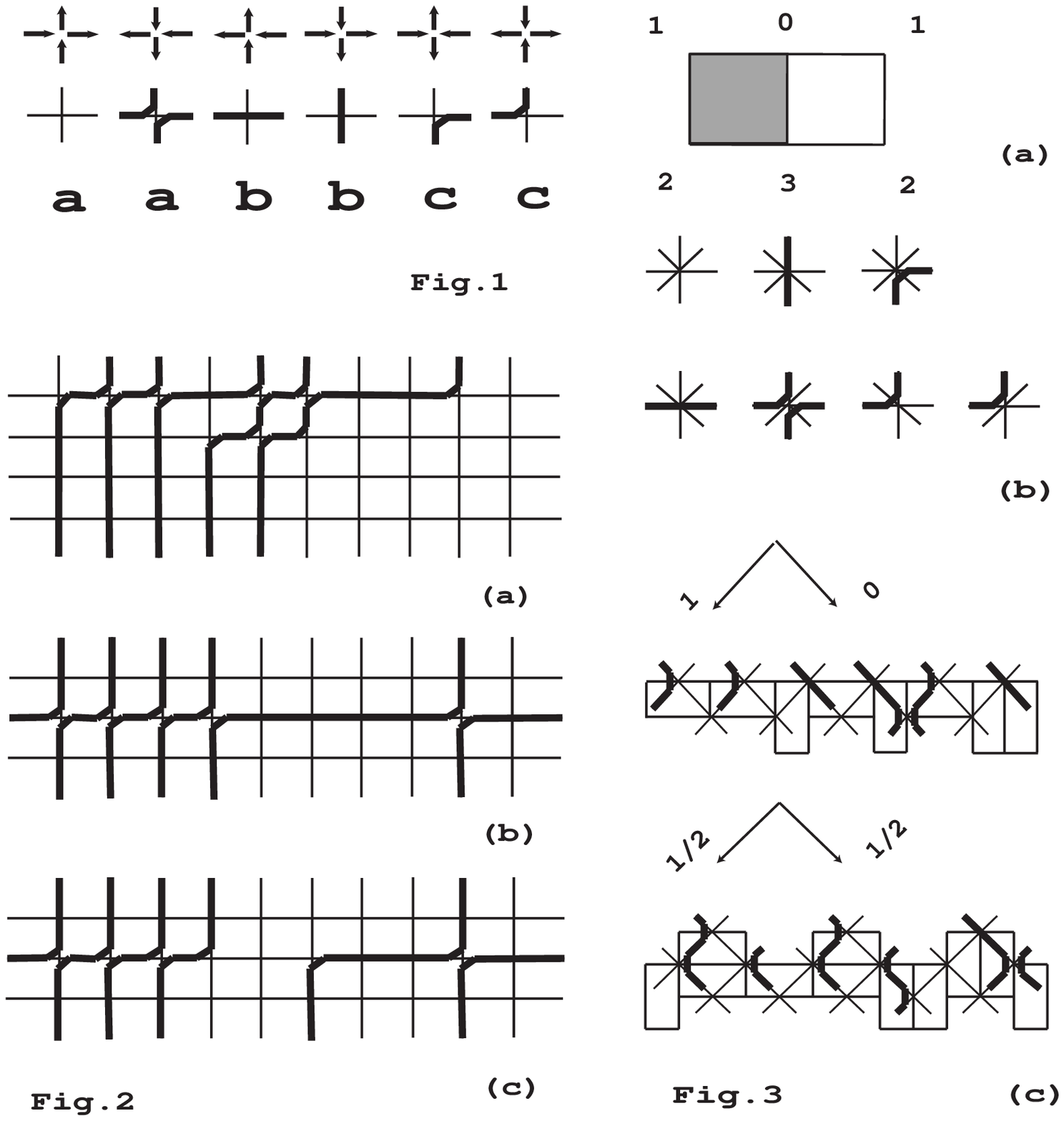}{-1.2in}{10.0in}
\end{figure}

\noindent
Figure Captions: 
\newline

\noindent
Fig.1: Six vertices, corresponding line configurations and their Boltzmann weights.
\newline

\noindent
Fig.2: Shift of lines introduced in the proof of Lemma 1 and Lemma 2.
\newline

\noindent
Fig.3: 
(a) The height function for domino configurations.
(b) Correspondence between vertices and domino tilings. The horizontal and the vertical lines are the lattice for the six-vertex model, while the lines rotated $\pm\pi/4$ are the edges of dominos.
(c) Two boundaries with constant tilt (0,0) but different line densities.
\newline





%

\end{document}